\documentclass[a4paper,14pt,psfig,epsfig,amsmath,amssymb]{article}
 
\usepackage{amssymb,epsfig}
\usepackage{amsfonts}
\usepackage{amsmath}
\usepackage{graphics}

\begin{document}
 
\begin{center} 
{\LARGE \bf
Chemical 
freeze-outs of strange and non-strange particles and residual chemical non-equilibrium
}

\vspace*{0.5cm}
{
K.~A.~Bugaev$^{1}$,  D.~R.~Oliinychenko$^{1,2}$, V.~V.~Sagun$^{1}$, A.~I.~Ivanytskyi$^{1}$, J. Cleymans$^{3}$, E. G. Nikonov$^{4}$ and G.~M.~ Zinovjev$^{1}$
}
\end{center} 

\vspace*{0.5cm}
 
{\hspace*{-0.45cm}$^{1}$ Bogolyubov Institute for Theoretical Physics, Metrologichna str. 14$^B$, Kiev 03680, Ukraine\\
 $^{2}$  FIAS, Goethe-University,  Ruth-Moufang Str. 1, 60438 Frankfurt upon Main, Germany\\
 $^{3}$ Department of Physics, University of Cape Town, Rondebosch 7701, South Africa\\
 $^{4}$ Laboratory for Information Technologies, JINR, Joliot-Curie str. 6, 141980 Dubna, Russia\\
 
}
 
\section{Abstract}
 
We propose an elaborate version of the hadron resonance gas model with the combined treatment of separate chemical  freeze-outs for strange and non-strange hadrons and with an additional $\gamma_{s}$ factor which accounts for the remaining  strange particle non-equilibration. Two sets of chemical freeze-outs parameters are connected by the conservation laws of entropy, baryonic charge, isospin projection and strangeness. The developed approach enables us to perform a high-quality fit of the hadron multiplicity ratios for AGS, SPS and RHIC energies with total  $\chi^2/dof \simeq$ 1.05. A special attention is paid to  a complete  description of  the  Strangeness Horn.  A  well-known $\bar p$,  $\bar \Lambda $ and $\bar \Xi$ selective suppression problem
is  also discussed.
 
\section{Introduction}
 
Relativistic A+A collisions are an important source of experimental information about the  QCD phase diagram and the strongly interacting matter properties.  The last stage of such collisions is traditionally analyzed within the statistical approach which gives us an excellent opportunity to reveal  the parameters of chemical freeze-out. This approach is based on the assumption of the  thermal equilibrium existence during  the last stage of reaction. Such an equilibrium can be reached due to intensive particle scattering. The stage of the system evolution when the inelastic reactions between hadrons stop is referred to as a  chemical freeze-out (FO). Particle yields are determined by the parameters of FO, namely by chemical potentials and temperature. This general picture is a basis of the Hadron Resonance Gas Model (HRGM) \cite{Thermal_model_review} which is the most successful one in describing the  hadronic  yields measured in heavy-ion experiments for energies from AGS to LHC. Despite a significant success of the HRGM in the  experimental data analysis there are a few  unresolved problems. In general they are related to the description of hadron yields which contain (anti)strange quarks. Especially the energy dependence of $K^{+}/\pi^{+}$ and $\Lambda/\pi^{-}$ ratios was out of high quality description. Excess of strange hadrons yields within the HRGM  led physical community to ponder over strangeness suppression.
The first receipt to resolve this problem was to introduce the strangeness suppression  factor $\gamma_{s}$ which should be fitted in order to describe the experimental data \cite{Rafelsky:gamma}.  However, such an approach is not supported by any underlying physical model and the physical meaning of $\gamma_{s}$ remains unclear \cite{KABAndronic:05,KABAndronic:09,KABugaev:Horn2013,Bugaev13,BugaevIndia}. In addition the strangeness suppression approach in its original form does not contain a hard-core repulsion between hadrons, while the latter is an important feature of the HRGM. A  significant role  
of the hard-core repulsion  was demonstrated once more  in Refs. \cite{KABugaev:Horn2013} where the global fit of hadron yield ratios was essentially improved (to $\chi^2/dof \simeq$ 1.2)  compared to all previous analyses.
 
The most advanced way to account for the hard-core repulsion between hadrons is to consider a hadron gas as a multi-component mixture of particles with different radii  \cite{KABugaev:Horn2013,Bugaev13,MultiComp:08,MultiComp:13}. Within  this approach   all baryons and mesons except for the kaons and the pions are endowed by the common hard-core radii $R_{b}$ and $R_{m}$,  respectively. At the same time the kaon and the  pion radii $R_{K}$ and $R_{\pi}$ are fitted independently in order to provide the best description of $K^{+}/\pi^{+}$ ratio \cite{KABugaev:Horn2013}. This  is an important 
finding since  the non-monotonic energy dependence of  $K^{+}/\pi^{+}$ ratio may indicate some  qualitative changes of the system properties and may  serve as a signal of the deconfinement onset. This is a reason why this ratio known as  the Strangeness Horn  is of  a special  interest. Note, that the multi-component approach substantially increased the  Strangeness Horn description quality, without spoiling the other ratios including 
$\Lambda/\pi^{-}$ one. However, even this advanced approach  does not reproduce the topmost  point of the Strangeness Horn indicating that the  data description is still not ideal. In order to resolve this problem in Ref.  \cite{Bugaev13} the $\gamma_{s}$ factor  was considered as a free parameter within the HRGM with 
multi-component repulsion.  
Although  the $\gamma_{s}$ data  fit   improves the Strangeness Horn description quality sizably,   it does not seem to be useful for  the description of other hadron multiplicities \cite{Bugaev13}.
 Furthermore, in contrast to the  claims  established on the low-quality fit \cite{Becattini:gammaHIC},  at low energies it was found \cite{Bugaev13} that within the error bars in heavy ion collisions there is an enhancement of strangeness and not a suppression. 
 
However,  the effect of apparent strangeness non-equilibration can be more successfully explained by the hypothesis of separate chemical FO for all strange hadrons. Since all the hadrons made of $u$ and $d$ quarks are under thermal equilibration whereas the  hadrons containing  $s$ quark are not,   then it is reasonable to assume two different FOs for these two kinds of particles. Following this  conclusion in Ref. \cite{Bugaev13, BugaevIndia} a separate strangeness FO (SFO) was introduced. Note, that according to  \cite{Bugaev13}  both FO and SFO parameters are  connected by the conservation laws of entropy, baryonic charge and isospin projection, while
the net strangeness is explicitly set  to zero at FO and at SFO. 
These conservation laws are  crucial elements of the concept of separate SFO developed in \cite{Bugaev13}
which allows  one to essentially reduce the number of  independent fitting parameters. 
Another principal element that differs the HRGM of  \cite{Bugaev13} from the ideal gas  treatment  used in  
\cite{BugaevIndia}  is the presence of multi-component hard-core repulsion.

Using the HRGM of    \cite{Bugaev13} it was possible to successfully describe
all hadron multiplicities measured in A+A collisions at AGS, SPS and RHIC energies   with $\chi^2/dof \simeq$ 1.06. The concept of separate SFO led to a systematic improvement of all experimental data description. However, the topmost  point of the Strangeness Horn again was not fitted even within the experimental error. Note, however,  that the general  description of $K^+/\pi^+$ ratio energy dependence was rather good except  for the upper point.
 
Since an introduction of the $\gamma_s$ factor demonstrated a remarkable description of all  points of the Strangeness Horn, whereas the separate SFO led to a systematic improvement of all hadron  yields description,  we decided to combine these elements in order to describe an experimental data with the highest possible quality. This ambitious task is the main aim of the present paper. In addition,  the problem of   residual strangeness non-equilibration  should also be  clarified due to its importance from the academic point of view.  
Evidently, the best tool for  such a  purpose is  the most successful  version of the HRGM, i.e. the HRGM  with the multi-component hadronic repulsion and SFO. 
As it will be shown below,  such an  approach makes it possible to describe  111 hadron yield ratios measured for 14 values of the center of mass collision energy $\sqrt{s_{NN}}$ in the interval from 2.7 GeV to  200 GeV with the highest quality ever achieved.
 
The paper is organized as follows. The basic features of the developed  model are outlined in  Section 3. In Section 4 we present and discuss the new fit of  hadronic multiplicity ratios   with two chemical freeze-outs and  $\gamma_{s}$ factor, while Section 5 contains our conclusions.

\section{Model description}

\begin{figure}[!]
\center{\includegraphics[width=80mm]{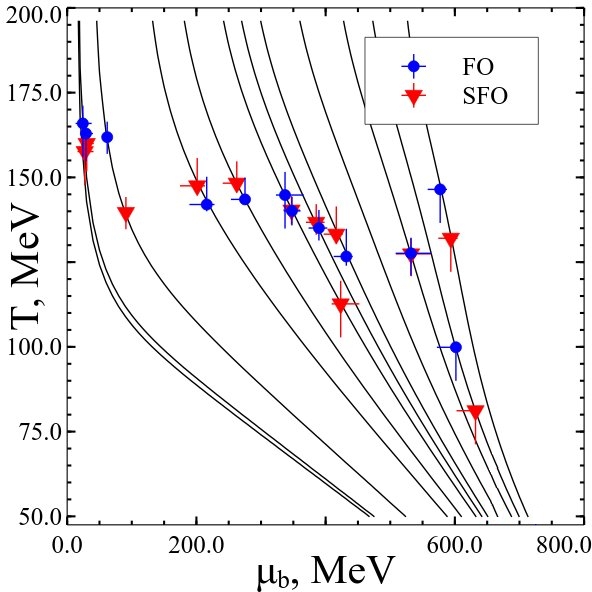}}
\caption{(Colour on-line) Chemical freeze-outs parameters in the model with two freeze-outs and with the $\gamma_{s}$  fit. Baryonic chemical potential dependence of the chemical freeze-out temperature for SFO (marked with triangles) and for FO (marked with circles). The solid  black curves correspond to the isentops $s/\rho_{B}=const$, on which the FO and the SFO points are located.}
\label{Fig:sagunI}
\end{figure}

\begin{figure}[!]
\begin{minipage}[h]{0.49\linewidth}
\center{\includegraphics[width=1.0\linewidth]{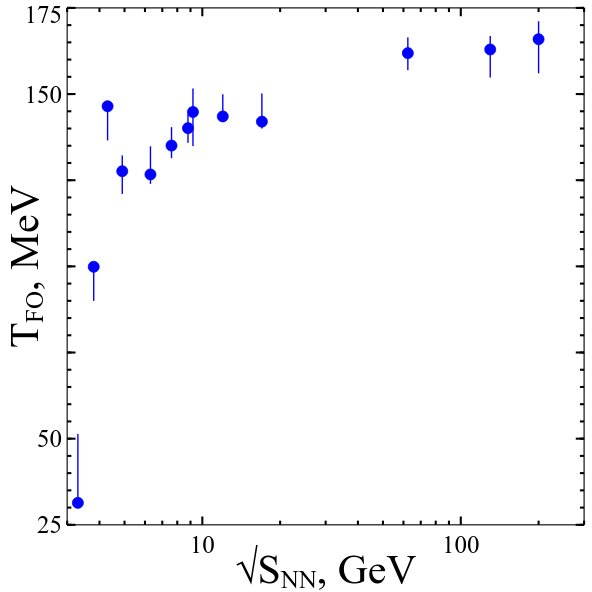}}
\end{minipage}
\hfill
\begin{minipage}[h]{0.49\linewidth}
\center{\includegraphics[width=1.0\linewidth]{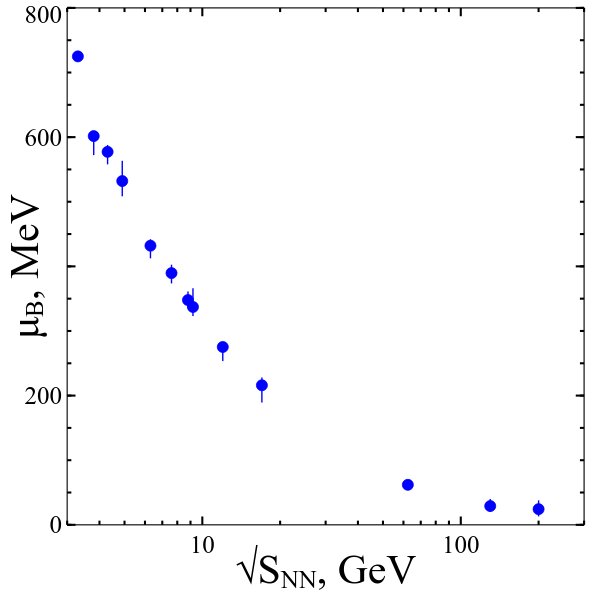}}
\end{minipage}
\caption{(Colour on-line) The behavior of the model parameters: chemical  freeze-out temperature T vs. $\sqrt{s_{NN}}$ (left panel) and the freeze-out baryonic chemical potential $\mu_{B}$ vs. $\sqrt{s_{NN}}$  (right panel).
}
\label{Fig:sagunII}
\end{figure}

In what follows we treat a hadronic system as a multi-component Boltzmann gas of hard spheres. The effects of quantum statistics are negligible for typical temperatures of the hadronic gas whereas the hard-core repulsion between the particles significantly affects a corresponding equation of state \cite{KABugaev:Horn2013,MultiComp:08}. The present model is dealing with the Grand Canonical treatment. Hence a thermodynamical state of   system under consideration is fixed by the volume $V$, the temperature $T$, the baryonic  
chemical potential $\mu_B$, the strange chemical potential $\mu_S$ and the chemical potential of the  isospin third component $\mu_{I3}$. These parameters control the pressure $p$ of the system. In addition they define the densities $n_i^K$ of corresponding charges $Q_i^K$ ($K\in\{B,S,I3\}$).  Introducing the symmetric matrix of the second virial coefficients $\mathcal{B}$ 
 with the elements $b_{ij} = \frac{2\pi}{3}(R_i+R_j)^3$, 
we can obtain the parametric equation of state of the present model in a compact form
\begin{eqnarray}\label{EqI}
\frac{p}{T} =  \sum_{i=1}^N \xi_i \,, ~~n^K_i =\frac{ Q_i^K{\xi_i}}{\textstyle  1+\frac{\xi^T {\cal B}\xi}{\sum\limits_{j=1}^N \xi_j}}, ~~\xi  = \left(
\begin{array}{c}
 \xi_1 \\
 \xi_2 \\
 ... \\
 \xi_N
\end{array}
\right) \,.
\end{eqnarray}
The equation of state is written in terms of the solutions $\xi_i$ of the following system
\begin{eqnarray}\label{EqII}
&&\hspace*{-4mm}\xi_i =\phi_i (T)\,   \exp\Biggl[ \frac{\mu_i}{T} - {\textstyle \sum\limits_{j=1}^N} 2\xi_j b_{ij}+{\xi^T{\cal B}\xi} {\textstyle \left[ \sum\limits_{j=1}^N\xi_j\right]^{-1}} \Biggr] \,, \quad \quad \\
&&\hspace*{-4mm}\phi_i (T)  = \frac{g_i}{(2\pi)^3}\int \exp\left(-\frac{\sqrt{k^2+m_i^2}}{T} \right)d^3k  \,.
%
\end{eqnarray}
It is worth to note, that quantities $T\xi_i$ have a meaning of $i^{th}$ sort of hadrons partial pressure. Each $i^{th}$ sort is characterized by  its  full chemical potential $\mu_i=Q_i^B\mu_i^B+Q_i^S\mu_i^S+Q_i^{I3}\mu_i^{I3}$, mass $m_i$ and degeneracy $g_i$. Function $\phi_i(T)$ denotes the  corresponding particle thermal density in case of ideal gas. Finally, the superscript $T$ here is the symbolic notation for operation of a column transposition which yields  a row of quantities $\xi_i$.
The obtained model parameters for two freeze-outs and their dependence on  the collision energy are shown in Figs. \ref{Fig:sagunI}-\ref{Fig:sagunIII}.
 
In order to  account for the possible  strangeness non-equilibration we introduce the  $\gamma_s$ factor in a conventional way by replacing $\phi_i$ in Eq. (\ref{EqII}) as
\begin{eqnarray} \label{eq:gamma_s}
\phi_i(T) \to \phi_i(T) \gamma_s^{s_i} \,,
\end{eqnarray}
where $s_i$ is a number of strange valence quarks plus number of strange  valence anti-quarks.
 
The principal  difference of the present  model from the  traditional approaches is that we employ an independent chemical FO of strange particles. Let us consider
this in some detail. The independent freeze-out of strangeness means that inelastic reactions (except for the decays) with  hadrons made of s quarks are switched off at the temperature $T_{SFO}$, the baryonic chemical potential $\mu_{B_{SFO}}$, the strange chemical potential
$\mu_{S_{SFO}}$, the isospin third projection chemical potential $\mu_{I3_{SFO}}$ and the three-dimensional emission volume $V_{SFO}$. In general case these parameters of SFO do not coincide with  the temperature $T_{SFO}$, the chemical potentials $\mu_{B_{SFO}}$, $\mu_{S_{SFO}}$,  
$\mu_{I3_{SFO}}$ and the volume $V_{SFO}$ which characterize the freeze-out of non-strange hadrons. The particle yields are given by the charge density $n^K_i$ in (1) and the corresponding 
volume 
at  FO and at  SFO. 
 
 At the first glance a model with independent SFO contains four extra fitting  parameters for each  energy value compared to the traditional  
approach (temperature, three chemical potentials and the volume at SFO instead of strangeness suppression/enhancement factor $\gamma_s$).  
However, this  is not the case due to the conservation laws. Indeed, since the  entropy, the baryonic charge and the isospin third projection are conserved,  then the  parameters of FO and SFO are connected by the following equations
\begin{eqnarray}
s_{FO} V_{FO} = s_{SFO} V_{SFO} \,, \label{ent_cons}\\
n^B_{FO} V_{FO} = n^B_{SFO} V_{SFO} \,, \label{B_cons}\\
n^{I_3}_{FO} V_{FO} = n^{I_3}_{SFO} V_{SFO} \,. \label{I3_cons}
\end{eqnarray}
The effective volumes can be excluded,  if these equations are rewritten as  
\begin{eqnarray}  
\label{Eq:FO_SFO1}
\frac{s}{n^B} \biggl|_{FO} = \frac{s}{n^B} \biggr|_{SFO} \,,  \quad  \frac{n^B}{n^{I_3}} \biggl|_{FO} = \frac{n^B}{n^{I_3}} \biggr|_{SFO}
%
 \,.   
\end{eqnarray}
Thus, the baryonic $\mu_{B_{SFO}}$ and the  isospin third projection $\mu_{I3_{SFO}}$ chemical potentials at SFO are now defined by Eqs. (\ref{Eq:FO_SFO1}). Note, that the strange chemical potentials $\mu_{S_{FO}}$ and $\mu_{S_{SFO}}$  are found from the condition of vanishing  net strangeness at FO and SFO,  respectively. Therefore, the concept of independent SFO leads to an appearance of one independently fitting parameter  $T_{SFO}$.
Hence, the independent fitting parameters are the following: the baryonic chemical potential $\mu_B$, the chemical potential of the third projection of isospin $\mu_{I3}$,  the chemical freeze-out  temperature for strange hadrons $T_{SFO}$, the chemical freeze-out  temperature for all non-strange hadrons $T_{FO}$ and the $\gamma_s$ factor. 
 
An inclusion  of  the width $\Gamma_i$ of  hadronic states  is an important element of the present model. It is due to the fact that the  thermodynamical properties of the hadronic system are sensitive to the width \cite{KABugaev:Horn2013, Bugaev13, Bugaev13new}. In order to account for the finite width of resonances we perform the usual modification of the thermal particle density $\phi_i$. Namely, we convolute the Boltzmann exponent under the integral over momentum with the normalized Breit-Wigner mass distribution. As a result, the modified thermal particle density of $i^{th}$ sort hadron acquires  the form
\begin{eqnarray}
\label{EqIII}
\int \exp\left(-\frac{\sqrt{k^2+m_i^2}}{T} \right)d^3k \rightarrow \frac{\int^{\infty}_{M_{0}} \frac{dx}{(x-m_{i})^{2}+\Gamma^{2}_{i}/4} \int \exp\left(-\frac{\sqrt{k^2+x^2}}{T} \right)d^3k }{\int^{\infty}_{M_{0}} \frac{dx}{(x-m_{i})^{2}+\Gamma^{2}_{i}/4}} \,. 
\end{eqnarray}
Here $m_i$ denotes the  mean mass of  hadron and $M_0$ stands for the threshold in the dominant decay channel. The main advantages of this approximation is a simplicity of  its realization and a clear way to account for the finite  width of hadrons.

The observed hadronic multiplicities  contain the thermal and decay contributions. For example, a large part of pions is produced by the decays of heavier hadrons. Therefore, the total multiplicity is obtained as a sum of thermal and decay multiplicities, exactly as it is done in a conventional model. However, writing the formula for final particle densities, we have to take into account that volumes of FO and SFO can be different:
\begin{eqnarray} \label{Eq:SFO_decays}
\frac{N^{fin}(X)}{V_{FO}} = \sum_{Y \in FO} BR(Y \to X) n^{th}(Y) + 
 \sum_{Y \in SFO} BR(Y \to X) n^{th}(Y) \frac{V_{SFO}}{V_{FO}} \,. 
\end{eqnarray}
Here the first term on the right hand side is due to decays after FO whereas the second one accounts for the strange resonances decayed after SFO. The factor $V_{SFO}/V_{FO}$ can be replaced by $n^B_{FO}/n^B_{SFO}$ due to baryonic  charge conservation. $BR(Y \to X)$ denotes the branching ratio  of the Y-th hadron decay into the X-th hadron, with the definition $BR(X \to X)$ = 1 used  for the sake of convenience. The input parameters of the present model (masses $m_i$, widths $\Gamma_i$, degeneracies $g_i$ and branching  ratios of all strong decays) were taken from the particle tables of the thermodynamical code THERMUS \cite{THERMUS}.

\section{Results}

{\bf Data sets  and fit procedure}. The present model is applied to fit the data. We take the ratios of particle multiplicities at midrapidity as the data points. In contrast to  fitting multiplicities themselves such an approach allows us  to cancel the possible experimental biases. In this paper we use the data set almost identical to Ref. \cite{Bugaev13}. At the AGS energies ($\sqrt{s_{NN}}=2.7-4.9$ AGeV or $E_{lab}=2-10.7$ AGeV) the data are available with a  good energy resolution above 2 AGeV. However, for the beam energies 2, 4, 6 and 8 AGeV only a few data points are available. They corresponds to the yields for pions \cite{AGS_pi1, AGS_pi2}, for protons \cite{AGS_p1,AGS_p2}, for kaons \cite{AGS_pi2} (expect for 2 AGeV). The integrated over $4\pi$ data are also available for $\Lambda$ hyperons \cite{AGS_L} and for $\Xi^-$ hyperons (for 6 AGeV only) \cite{AGS_Kas}. However, as was argued in Ref. \cite{KABAndronic:05}, the data for $\Lambda$ and $\Xi^-$ should be recalculated for midrapidity. Therefore, instead of raw experimental data we used the corrected values from \cite{KABAndronic:05}. Next comes the data set at the highest AGS energy ($\sqrt{s_{NN}}=4.9$ AGeV or $E_{lab}=10.7$ AGeV). Similarly to \cite{KABugaev:Horn2013}, here  we analyzed  only  the  NA49  mid-rapidity data   \cite{KABNA49:17a,KABNA49:17b,KABNA49:17Ha,KABNA49:17Hb,KABNA49:17Hc,KABNA49:17phi}. Since  the RHIC high energy  data of different collaborations agree with each other, we  analyzed  the STAR results  for  $\sqrt{s_{NN}}= 9.2$ GeV \cite{KABstar:9.2}, $\sqrt{s_{NN}}= 62.4$ GeV \cite{KABstar:62a}, $\sqrt{s_{NN}}= 130$ GeV \cite{KABstar:130a,KABstar:130b,KABstar:130c,KABstar:200a} and  200 GeV \cite{KABstar:200a,KABstar:200b,KABstar:200c}.  
 
The criterion to define  the fitting parameters of the present model is a minimization of $\chi^2=\sum_i\frac{(r_i^{ther}-r_i^{exp})^2}{\sigma^2_i}$, where $r_i^{theor}$ and $r_i^{exp}$ are, respectively,  the  theoretical and the experimental  values of particle yields ratios, $\sigma_i$ stands for the corresponding experimental error and a  summation is performed over all available experimental points.
 
{\bf  Combined fit with SFO and $\gamma_{s}$ factor}. Recently performed comprehensive data analysis \cite{Bugaev13} for  two alternative  approaches, i.e  the first one with $\gamma_{s}$ as a free parameter and the second one with separate FO and SFO, showed  the advantages and disadvantages of both methods. Thus, the $\gamma_{s}$ fit provides one  with an opportunity to noticeably improve the Strangeness Horn description with $\chi^2/dof=3.3/14$, comparably to the previous result $\chi^2/dof=7.5/14$ \cite{KABugaev:Horn2013}, but there are only slight improvements of the ratios with strange baryons (global $\chi^2/dof: 1.16 \rightarrow 1.15$). The obtained results for the SFO approach demonstrate a nice fit quality for the most problematic ratios for the HRGM, especially for   $\bar p/\pi^-$, $\bar \Lambda/\Lambda$,  $\bar \Xi^-/\Xi^-$ and  $\bar \Omega/\Omega$. Although the overall $\chi^2/dof \simeq 1.06$ is notably better than with the  $\gamma_{s}$ factor \cite{KABugaev:Horn2013, Bugaev13}, but the highest point fitting of the Horn got worse. These results  led  us to an idea to investigate the combination of these two approaches in order to get the high-quality Strangeness Horn description without spoiling the quality of other particle ratios.

\begin{figure}[!]
\center{\includegraphics[width=80mm]{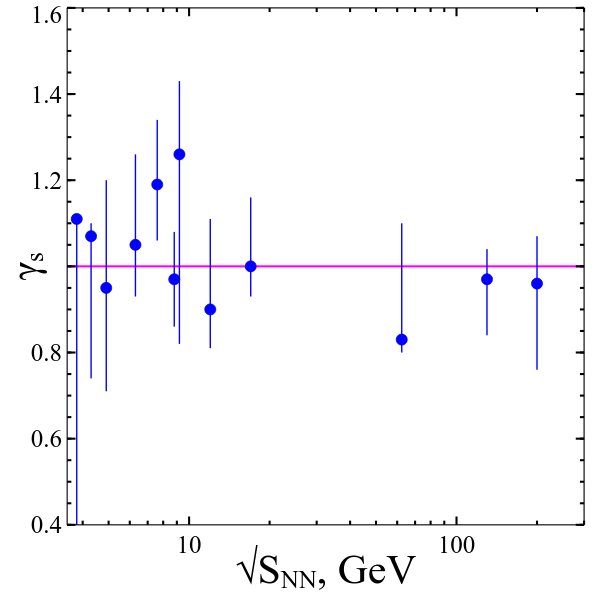}}
\caption{(Colour on-line) $\sqrt{s_{NN}}$ dependence of the $\gamma_s$ factor in the model with two freeze-outs and the  $\gamma_s$ fit.}
\label{Fig:sagunIII}
\end{figure}
 
For 14 values of collision energy $\sqrt{s_{NN}} = $  2.7, 3.3, 3.8, 4.3, 4.9, 6.3, 7.6, 8.8, 9.2, 12, 17, 62.4, 130, 200 GeV the best description with two separate freeze-outs and the $\gamma_s$ fit gives  $\chi^2/dof$ = 42.96/41 $\simeq$ 1.05, which is only a very slight improvement compared to the previously obtained  results $\chi^2/dof$ =58.5/55$\simeq$ 1.06 found for two freeze-outs (strange and all other particles), which are connected by the conservation laws. Note, however, that the value of $\chi^2$ itself, not divided by number of degrees of freedom,  has improved notably, although the deviation of the $\gamma_s$ factor from 
1  does not exceed 27 \% (see Fig. \ref{Fig:sagunIII}).
These findings  motivate us to study what ratios and at what energies  are  improved.
 
As  we  mentioned earlier, at each collision energy there are five independent fitting parameters in the considered model with the simultaneous SFO and the $\gamma_{s}$ fit, while for some collision energies the number of  experimental  ratios  
is lower or  equal to the   number of parameters. For example, for the energies $\sqrt{s_{NN}}$ =2.7, 3.3, 3.8, 4.3, 9.2, 62.4 GeV the number of available ratios is small (4, 5, 5, 5, 5, 5, respectively) from which only kaons and $\Lambda$ contain strange quarks. Therefore, for these energies we obtained a perfect data description 
since we had  to solve the above equations.  As a result for these energies 
the relative deviation of the fit is  almost  a zero, but   it gives us  somewhat larger uncertainties for  the fitting parameters.
 
\begin{figure}[!]
\center{\includegraphics[width=80mm]{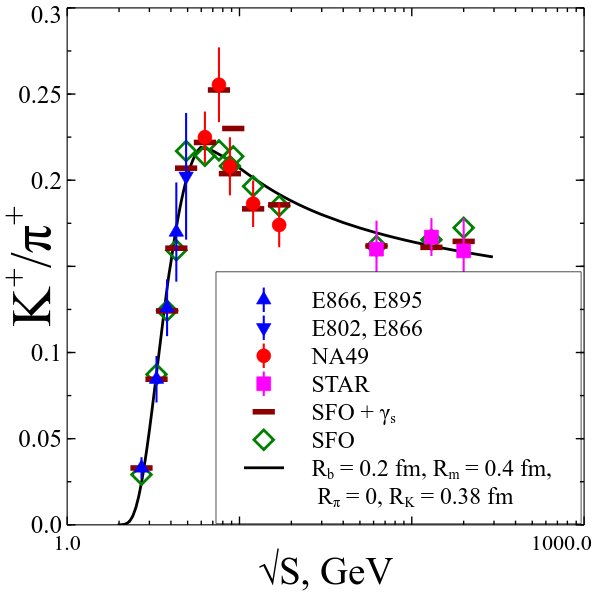}}
\caption{(Colour on-line)  $\sqrt{s_{NN}}$  dependences of  $K^+/\pi^+$ ratio. The solid line corresponds to the results of \cite{KABugaev:Horn2013}. Horizontal bars correspond to the present  model with SFO+$\gamma_s$ fit, while the  diamonds  correspond  to the  results  previously obtained for  SFO \cite{Bugaev13}.
}
\label{Fig:sagunIV}
\end{figure}
 
\begin{figure}[!]
\begin{minipage}[h]{0.49\linewidth}
\center{\includegraphics[width=1.0\linewidth]{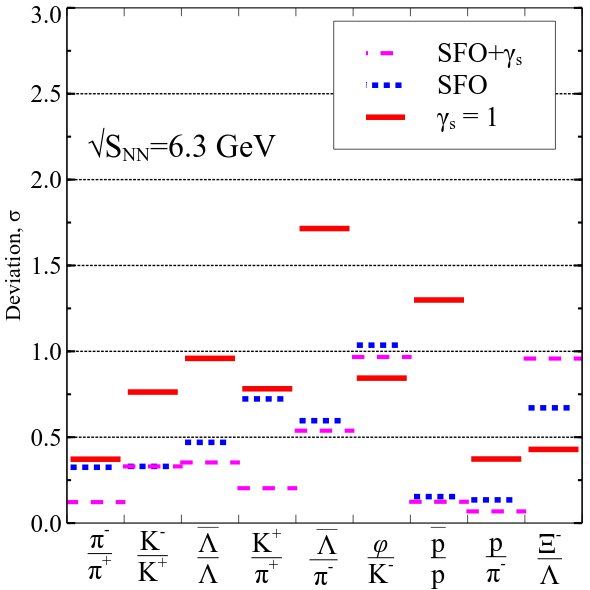}}
\end{minipage}
\hfill
\begin{minipage}[h]{0.49\linewidth}
\center{\includegraphics[width=1.0\linewidth]{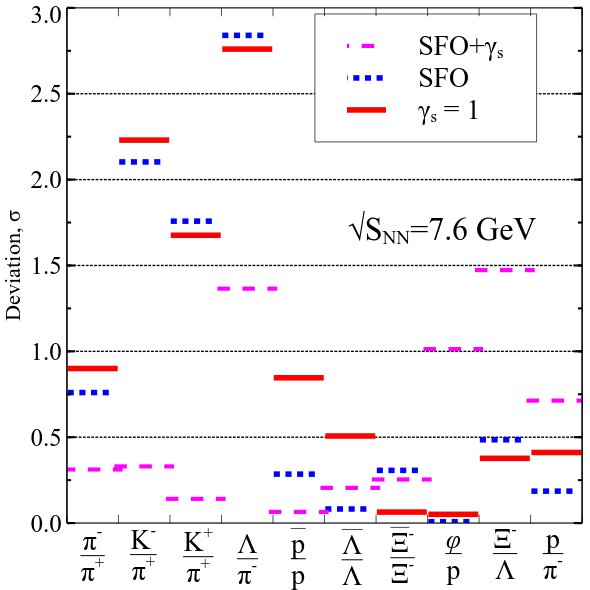}}
\end{minipage}
\begin{minipage}[h]{0.49\linewidth}
\center{\includegraphics[width=1.0\linewidth]{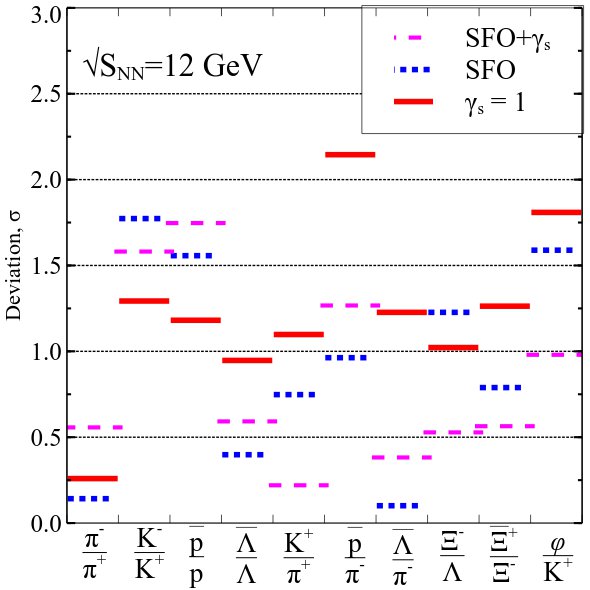}}
\end{minipage}
\hfill
\begin{minipage}[h]{0.49\linewidth}
\center{\includegraphics[width=1.0\linewidth]{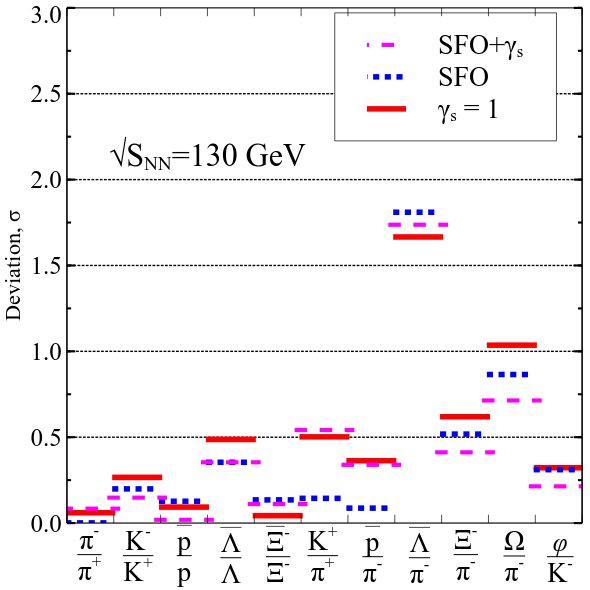}}  
\end{minipage}
\caption{(Colour on-line) Relative deviation of the theoretical description of ratios from the experimental value in units of the experimental error $\sigma$. Particle ratios vs. the modulus of relative deviation ($\frac{|r^{theor} - r^{exp}|}{\sigma^{exp}}$) for  $\sqrt{s_{NN}} = $ 6.3, 7.6, 12 and 130 GeV are shown. Solid lines correspond to the model with a single FO of all hadrons and $\gamma_s =1$, blue dotted lines correspond to the model with SFO. The results of a   model with  a combined  fit with SFO and  $\gamma_s$    are  highlighted by magenta dashed lines.
}
\label{Fig:sagunV}
\end{figure}
 
\begin{figure}[!]
\begin{minipage}[h]{0.49\linewidth}
\center{\includegraphics[width=1.0\linewidth]{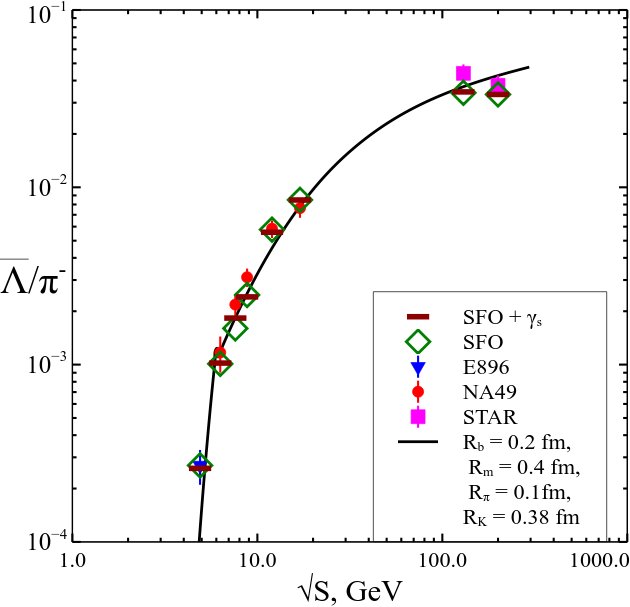}}
\end{minipage}
\hfill
\begin{minipage}[h]{0.49\linewidth}
\center{\includegraphics[width=1.0\linewidth]{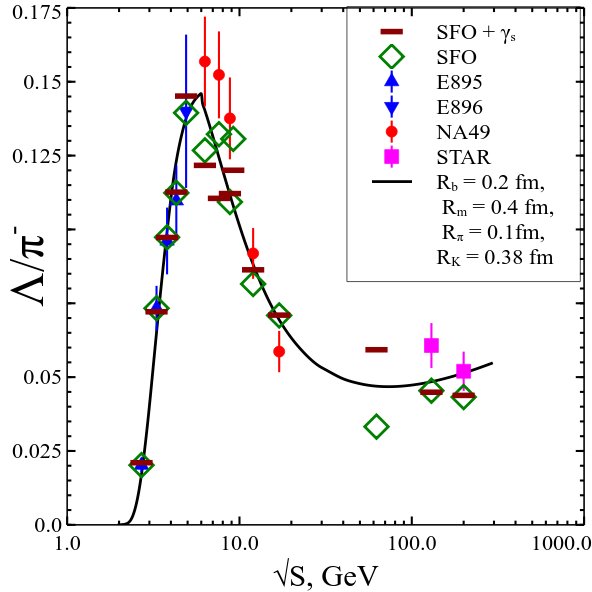}}
\end{minipage}
\caption{(Colour on-line)  $\sqrt{s_{NN}}$  dependences of  $\bar \Lambda/\pi^-$ (left panel) and $\Lambda/\pi^-$ (right panel) ratios. The solid line correspond to the results of  \cite{KABugaev:Horn2013}.  Horizontal bars correspond to SFO+$\gamma_s$ model, while green diamonds  correspond  to the previously obtained results  for the  SFO model \cite{Bugaev13}.
}
\label{Fig:sagunVI}
\end{figure}
 
For the   energies $\sqrt{s_{NN}} = $ 17, 130 GeV  we observed that the resulting fit quality became better compared to the  work \cite{Bugaev13}. The most significant improvements correspond to the collision energies  $\sqrt{s_{NN}} = $ 6.3, 7.6, and 12 GeV, that are plotted in Fig. \ref{Fig:sagunV}. Fig.  \ref{Fig:sagunV} demonstrates very nice fit quality, especially for such  traditionally  problematic ratios as $K^+/\pi^+$, $\pi^-/\pi^+$, $\bar \Lambda/\pi^-$ and $\varphi/K^+$. For $\sqrt{s_{NN}} = $ 7.6 GeV  the seven ratios out of ten  are improved.  
 
A special attention in our consideration  was paid to the Strangeness Horn, i.e. to   $K^+/\pi^+$ ratio. Another reason for a through   study of the Strangeness Horn is a traditional   problem of the HRGM to fit  it. As one can see  from Fig. \ref{Fig:sagunIV}, the remarkable  $K^+/\pi^+$  fit improvement  for $\sqrt{s_{NN}} = $ 2.7, 3.3, 4.3, 4.9, 6.3, 7.6, 12 GeV justifies the usage of the present  model. Quantitatively, we found that $\chi^2/dof$ improvement due to SFO+$\gamma_s$ introduction is  $\chi^2/dof$=1.5/14, i.e. even better than it was done in \cite{Bugaev13} with $\chi^2/dof$=3.3/14 for the $\gamma_s$ fitting approach and  $\chi^2/dof$=6.3/14 for   SFO and  $\gamma_s=1$.
 
In addition, in Fig. \ref{Fig:sagunVI} we give   the $\Lambda/ \pi^-$ and $\bar \Lambda/ \pi^-$  ratios to show that two separate freeze-outs inclusion  with the $\gamma_s$ fit still  does not improve these ratios. The $\Lambda/ \pi^-$  fit quality, for instance, is  ($\chi^2/dof$=10/8).  Hence, up to now  the best fit of  the $\Lambda/ \pi^-$ ratio  was obtained within  the SFO approach with $\gamma_s=1$. As it  was mentioned in \cite{KABAndronic:05,KABAndronic:09,KABugaev:Horn2013}
   a too  slow decrease of model results for $\Lambda/ \pi^-$ ratio   compared to the experimental data is typical for almost all statistical models. Evidently, the too steep rise in $\Lambda/ \pi^-$ behavior is a consequence of the $ \bar \Lambda$ anomaly \cite{KABAndronic:05,AGS_L3}. Similar results are reported in Refs. \cite{KABugaev:Becattini, KABugaev:Becattini13, KABugaev:Stachel}  as  the $\bar p$,  $\bar \Lambda $ and $\bar \Xi$ selective suppression. Since even an  introduction of the separate strangeness freeze-out with the strangeness enhancement  factor does not allow us  to better describe   these  ratios, we believe that  there is  an  unclarified physical reason which is responsible  for   it.

Within the present model  we also found a selective improvement and a certain degradation of the fit quality of various ratios for different collision energies. For instance, the  $\pi^-/\pi^+$ ratio is slightly  increased  for $\sqrt{s_{NN}} = $   6.3 and 7.6 GeV, but the situation drastically changes for $\sqrt{s_{NN}} = $ 12 GeV. The same tendency  is typical for  $\bar p/p$. On the contrary, for  $\bar \Xi^-/\Lambda$ ratio there  is a noticeably  worse data description within   SFO+$\gamma_s$ approach at $\sqrt{s_{NN}} = $  6.3, 7.6 GeV, but for $\sqrt{s_{NN}} = $ 12 GeV the fit quality is sizably better compared  to  all previous approaches. Thus, within the present model we  reveal a noticeable change in the trend  of   some ratios  at  $\sqrt{s_{NN}} = $  7.6-12 GeV .
 
\section{Conclusions}
 
We have performed an elaborate  fit  of  the data measured at  AGS, SPS and RHIC energies within the multi-component hadron resonance gas model. The suggested  approach to  separately treat   the  freeze-outs of strange and non-strange  hadrons with the  simultaneous $\gamma_s$ fitting  gives  rise for the top-notch Strangeness Horn description with $\chi^2/dof$=1.5/14. The developed model   clearly  demonstrates  that the successful fit of  hadronic multiplicities  includes all the  advantages of these two approaches  discussed in  \cite{Bugaev13}.
As a result  for $\sqrt{s_{NN}} = $ 6.3, 7.6, 12, 130 GeV we found a significant data fit  quality improvement. The achieved  total value of $\chi^2/dof$ is  $42.96/41  \simeq$ 1.05, while the  $\gamma_s$ values are  consistent with the conclusion  $\gamma_s \simeq 1$ (within the error bars). A possible exception is the topmost  point of the Strangeness Horn, at which the mean value of the strangeness 
enhancement factor is $\gamma_s \simeq 1.27 $, but with the large error bars. 
 In addition, the description of ratios containing the   non-strange particles, especially such  as $\pi^-/\pi^+$ and $\bar p/p$,  gets better compared  to  previously reported  results \cite{KABugaev:Horn2013, Bugaev13}.  At the same time the lack of  available  data at 
$\sqrt{s_{NN}}$ =2.7, 3.3, 3.8, 4.3, 9.2, 62.4 GeV forced us to solve the corresponding equations 
which in combination with the  large experimental  error bars led to rather large uncertainties of  the fitting parameters.

From  a significant improvement of the data description we conclude that the concept of separate chemical freeze-out  of strange particles is an essential  part  of  heavy ion collision  phenomenology which should be taken into account in further studies of strongly interacting matter properties.  However,  the remaining  problem with  $\bar p$,  $\bar \Lambda$,  $\bar \Xi$ ratios led us to a conclusion  that there is an  unclarified physical reason which is responsible  for them. 
The  residual non-equilibration of strange particles found here  seems to be  weak and, perhaps, the better experimental data will help us to reduce it further. 

The obtained  description of  the hadron multiplicity ratios reached  the   highest  quality ever achieved   and this fact demonstrates  that the  suggested approach is almost a precise tool to elucidate the thermodynamics properties of hadron matter at two chemical freeze-outs. The fresh  illustrations  to this statement can be found in \cite{Bugaev13new}.  However, to get  more reliable conclusions from this approach we need 
more experimental data with  an essentially  higher accuracy. 

\vspace*{3mm}

\noindent
{\bf Acknowledgments.}  We would like to thank A. Andronic for  providing an access to well-structured experimental data.
The authors are thankful to  I. N. Mishustin, D. H. Rischke  and L. M. Satarov for valuable comments.
K.A.B., A.I.I.  and G.M.Z.  acknowledge  a  support 
of  the Fundamental Research State Fund of Ukraine, Project No F58/04.
K.A.B.   acknowledges  also  a partial support provided by the Helmholtz 
International Center for FAIR within the framework of the LOEWE 
program launched by the State of Hesse.

\end{document}